\documentclass[conference]{IEEEtran}
\IEEEoverridecommandlockouts
\usepackage{cite}
\usepackage{amsmath,amssymb,amsfonts}
\usepackage{algorithmic}
\usepackage{graphicx}
\usepackage{textcomp}
\usepackage{xcolor}
\def\BibTeX{{\rm B\kern-.05em{\sc i\kern-.025em b}\kern-.08em
    T\kern-.1667em\lower.7ex\hbox{E}\kern-.125emX}}

\usepackage{bm}
\usepackage[boxed]{algorithm2e}
\usepackage{hyperref}

\newtheorem{theorem}{Theorem}[section]
\newtheorem{definition}{Definition}[section]
\newtheorem{assumption}{Assumption}
\SetAlCapSkip{.75em}

\begin{document}

\title{Network Recovery from Unlabeled Noisy Samples
\thanks{This research was partially supported by ARO award W911NF1810237.}
}

\author{\IEEEauthorblockN{Nathaniel Josephs, Wenrui Li, and Eric. D. Kolaczyk}
\IEEEauthorblockA{\textit{Dep. of Math. and Stat.} \\
\textit{Boston University}\\
Boston, MA, 02215, USA \\
\{njosephs, wenruili, kolaczyk\}@bu.edu}
}

\maketitle

\begin{abstract}
There is a growing literature on the statistical analysis of multiple networks in which the network is the fundamental data object.
However, most of this work requires networks on a shared set of labeled vertices.
In this work, we consider the question of recovering a parent network based on noisy unlabeled samples.
We identify a specific regime in the noisy network literature for recovery that is asymptotically unbiased and computationally tractable based on a three-stage recovery procedure:
first, we align the networks via a sequential pairwise graph matching procedure; next, we compute the sample average of the aligned networks; finally, we obtain an estimate of the parent by thresholding the sample average.
Previous work on multiple unlabeled networks
is only possible for trivial networks due to the complexity of brute-force computations.
\end{abstract}


\section{Introduction}

Networks are widely used across various scientific disciplines and are the preeminent method for representing relational data.
The study of individual networks is well established, but technological advances are making data sets with multiple networks increasingly common.
Consequently, there is a growing literature on the statistical analysis of multiple network data that includes solutions for estimating a population network based on noisy realizations \cite{le2018estimating, wang2019common}, modeling distributions over networks \cite{durante2017nonparametric}, and averaging \cite{jain2016statistical, ginestet2017hypothesis},
hypothesis testing \cite{ghoshdastidar2020two, chen2020spectral}, and classification \cite{relion2019network, josephs2020bayesian} for collections of networks.
There is also a large literature on graph signal processing \cite{ortega2018graph} with solutions for combining \cite{egilmez2016optimization}, clustering \cite{tang2009clustering}, and classifying \cite{menoret2017evaluating} multiple graphs.
Work in this area, including all of the aforementioned work, generally assumes networks with labeled nodes.

Unlabeled networks, on the other hand, have been less studied, but are now emerging as relevant objects in many areas including differential privacy and anonymized networks \cite{narayanan2009anonymizing, pedarsani2011privacy, rossi2015k}, registration of brain networks due to mismeasurement (e.g, misalignment of brain regions due to batch effects of brain scans) \cite{lyzinski2014seeded} or to elastic shapes \cite{guo2020statistical}, and active learning of unlabeled nodes \cite{kajdanowicz2016learning}.
Therefore, it is important to try to obtain similar tools for studying multiple unlabeled networks.

In this work, we turn to the growing literature on noisy networks.
There, a variant of the traditional `signal-plus-noise' framework has been found useful for quantifying the uncertainty of various network characteristics due to observing a noisy version of a true underlying network.
Specifically, an underlying network $\bm A$ may have some characteristic $\eta(\bm A)$ that is of interest, and uncertainty in either the empirical version $\eta(\tilde{\bm A})$ and/or improved estimators are studied based on a noisy observation $\tilde{\bm A}$.
This framework has been studied when the network characteristics of interest are subgraph counts estimated from a single noisy network \cite{balachandran2017propagation}, as well as subgraph densities \cite{chang2020estimation} and branching factors \cite{li2020estimation} which require several noisy replicates.
Herein, we assume $m \geq 2$ noisy networks are observed, without node labels, and our goal is to recover the entire underlying network $\bm A$, i.e $\eta(\cdot)$ is the identity.

In solving the problem of network recovery given a sample of unlabeled noisy networks, we also introduce an efficient computation of the network average in our setting.
This contribution is of independent interest, since previous work on computing averages for unlabeled networks is limited by the computational complexity of their solutions.
For instance, in \cite{kolaczyk2020averages}, by characterizing the geometry of the space of unlabeled networks and associating with that space a Procrustean distance, the authors are led to a notion of nonparametric average (the Fr\'echet mean) for which brute-force computation yields an $O((n!)^m)$ algorithm, where $m$ is the number of networks and $n$ is the number of nodes.
Iterative approaches have been proposed based on graph matching ideas for computing the sample Fr\'echet mean of unlabeled networks \cite{jain2016statistical, guo2019quotient, calissano2020populations}, but these are $O(m(n!))$ and thus still too expensive for large networks.

The remainder of this paper is organized as follows.
In Section \ref{sec:noisy}, we discuss our problem setting in which multiple noisy networks are observed and we make a connection between the `signal-plus-noise' model and the correlated Erd\H{o}s-R\'{e}nyi model.
We introduce a three-staged approach for estimating the parent network given a collection of unlabeled noisy networks in Section \ref{sec:avg} and discuss the theoretical properties of the resulting estimator in Section \ref{sec:theory}.
A simulation study is presented in Section \ref{sec:simulation} and we conclude with a discussion of possible future directions for this work in Section \ref{sec:conclusion}.

\section{Noisy Networks}\label{sec:noisy}

In the noisy network literature, one is assumed to have a noisy version $\tilde{\bm A}$ of a true underlying network $\bm A$, where the latter is unobserved.
Interest frequently is in estimating a network characteristic $\eta(\bm A)$, for which the default choice is the plug-in estimate $\eta(\tilde{\bm A})$.
Typically, one assumes that
\begin{align}
    \mathbb{P}(\tilde{A}_{uv} = 1~|~A_{uv}) = \begin{cases}
        1-\beta_{uv} &\text{if}~A_{uv} = 1 \\
        \alpha_{uv} &\text{if}~A_{uv} = 0
    \end{cases} \enskip , \label{eq:noise}
\end{align}
where often these error rates are assumed to be constant, i.e $\alpha_{uv} \equiv \alpha$ and $\beta_{uv} \equiv \beta$ for $\alpha, \beta \in [0, 1]$.
Furthermore, in all of the settings in which multiple replicates are observed, the node labels are implicitly assumed to be known.

Throughout, we assume that $m \geq 2$ noisy networks are observed following \eqref{eq:noise}, but their node labels are unobserved, and we are interested in recovering the entire true underlying network $\bm A$.
To do so, we make a connection between the `signal-plus-noise' model in \eqref{eq:noise} and the correlated Erd\H{o}s-R\'{e}nyi model.

\subsection{Correlated Erd\H{o}s-R\'{e}nyi Model}

The correlated Erd\H{o}s-R\'{e}nyi model was introduced by \cite{pedarsani2011privacy} and has been used extensively in the context of graph matching \cite{yartseva2013performance, lyzinski2014seeded, dai2019analysis, cullina2019partial, barak2019nearly, ding2020efficient, hall2020partial, wu2021settling, ganassali2021impossibility} as well as testing edge correlation between two unlabeled graphs \cite{wu2020testing}.
This popular model is defined as follows.

\begin{definition}[Correlated Erd\H{o}s-R\'{e}nyi Model] \label{def:corrER}
Given an integer $n$ and $q, s \in [0, 1]$, let $\tilde{\bm A}^{(1)}$ and $\tilde {\bm A}^{(2)}$ denote the adjacency matrices of two Erd\H{o}s-R\'{e}nyi random graphs on the same vertex set $[n]=\{1,\ldots,n\}$ with probability $q \in [0, 1]$.
Let ${\pi}^* : [n] \rightarrow [n]$ denote a latent permutation.
We assume that conditional on $\tilde {\bm A}^{(1)}$, for all $u < v$, $\tilde A^{(2)}_{{\pi}^*(u){\pi}^*(v)}$ are independent and distributed as
\begin{align*}
    \tilde A^{(2)}_{{\pi}^*(u),{\pi}^*(v)} \sim \begin{cases}
        \text{Bern}(s) &\text{if}~\tilde A^{(1)}_{u,v} = 1 \\
        \text{Bern}\Big(\frac{q(1-s)}{1-q}\Big) &\text{if}~\tilde A^{(1)}_{u,v} = 0
    \end{cases} \enskip .
\end{align*}
\end{definition}

It follows from Definition \ref{def:corrER} that marginally we have $\tilde{\bm A}^{(2)} \sim \text{ER}(n, q)$, since for all $u < v$, we have
\begin{align*}
    \mathbb{P}\big(\tilde A^{(2)}_{{\pi}^*(u),{\pi}^*(v)} = 1\big) &= s \cdot \mathbb{P}(\tilde A^{(1)}_{u, v} = 1) \\
    &\ \ \ + \frac{q(1-s)}{1-q}\cdot\mathbb{P}(\tilde A^{(1)}_{u, v} = 0) \\
    &= s\cdot q + \frac{q(1-s)}{1-q}\cdot (1-q) \\
    &= q \enskip .
\end{align*}

Note that the original correlated Erd\H os-R\'enyi model in Definition \ref{def:corrER} can be seen as a particular case of the `signal-plus-noise' model in \eqref{eq:noise} in which the true underlying network is Erd\H os-R\'enyi, $\alpha = 0$ , and the node labels of the noisy replicates are permuted after sampling.
To see the connection between these models, let $\tilde{\bm A}^{(1)}$ and $\tilde{\bm A}^{(2)}$ be drawn from \eqref{eq:noise}, whose entries above the diagonal are independent conditional on some $\bm A \sim ER(n, p)$.
Suppose the nodes in $\tilde{\bm A}^{(2)}$ are then permuted by a latent permutation $\pi^*$.
It follows that $\tilde{\bm A}^{(1)}$ and $\tilde{\bm A}^{(2)}$ are samples from a correlated Erd\H os-R\'enyi model by letting $q$ and $s$ satisfy
\begin{align}
    \begin{split}
        q &= p(1-\beta)+(1-p)\alpha \enskip , \\
        s &= \frac{p(1-\beta)^2+(1-p)\alpha^2}{p(1-\beta)+(1-p)\alpha}  \enskip .
    \end{split} \label{eq:qs}
\end{align}
When $\alpha = 0$, this reduces to the equivalent generative formulation of the correlated Erd\H os-R\'enyi model in Definition \ref{def:corrER} with $q = ps$ and $s = 1 - \beta$.

This insight -- that $q$ and $s$ can be defined to include $\alpha > 0$ -- allows us to view the correlated Erd\H{o}s-R\'{e}nyi graph model as a generative formulation of two networks being noisy samples of some (unknown) true underlying Erd\H{o}s-R\'{e}nyi network.
These samples have the same natural interpretation as noisy replicates in the `signal-plus-noise' model, which are noisy unlabeled observations of a true underlying network.
Furthermore, $\alpha > 0$ enriches the correlated Erd\H{o}s-R\'{e}nyi model by allowing for the common scenario of observing a spurious edge while simultaneously retaining the efficient solutions designed for graph matching in the correlated Erd\H{o}s-R\'{e}nyi setting.
Finally, by taking this generative perspective,
it becomes clear that a collection of noisy networks from \eqref{eq:noise} has the property that each pair comes from a correlated Erd\H{o}s-R\'{e}nyi model with the same parent if that parent is assumed to be Erd\H{o}s-R\'{e}nyi.

\section{Recovery}\label{sec:avg}

In this section, we propose a three-staged approach for recovering an underlying network based on multiple unlabeled noisy networks.
The first stage involves aligning the unlabeled noisy networks via a sequential pairwise graph matching procedure.
This procedure yields estimates of the latent permutations.
In the second stage, we align the networks via their estimated permutations and compute the sample average
\begin{align}\label{eq:avg}
    \bar{\bm A} = \frac{1}{m}\sum_{i=1}^m \hat{\bm A}^{(i)} \enskip ,
\end{align}
where $\hat{\bm{A}}^{(i)}$ denotes the adjacency matrix for $\tilde{\bm A}^{(i)}$ that has been aligned to $\tilde{\bm A}^{(1)} \equiv \hat{\bm A}^{(1)}$.
We obtain our estimate of the parent network in the final stage by binarizing each entry of the sample average from \eqref{eq:avg} based on some threshold $w$:
\begin{align}\label{eq:est}
    \hat{A}_{uv} = \begin{cases}
        1 & \text{if}~\bar{A}_{uv} > w \\
        0 & \text{if}~\bar{A}_{uv} < w
    \end{cases} \enskip .
\end{align}

In the remainder of this section, we provide details for the sequential graph matching procedure and we discuss its computational cost as well as two improvements beyond the pairwise approach in the case that $m > 2$.

\subsection{Multiple Graph Matching}

Given a collection of unlabeled networks $\tilde{\bm{A}}^{(1)}, \ldots, \tilde{\bm{A}}^{(m)}$ sampled independently from \eqref{eq:noise} conditional on some fixed but unknown $\bm A$, we estimate the latent permutations in order to find a common alignment.
We do this by leveraging results from the graph matching literature \cite{conte2004thirty, lyzinski2015graph, yan2016short}.
The graph matching problem is to find an optimal alignment, or an isomorphism depending on if an exact solution is possible, given two unlabeled networks.

There have been several graph matching techniques specfically designed for correlated Erd\H{o}s-R\'{e}nyi networks including seeded graph matching \cite{lyzinski2014seeded}, percolation graph matching \cite{yartseva2013performance, barak2019nearly}, canonical labeling \cite{dai2019analysis}, and $k$-core alignment \cite{cullina2019partial}.
Herein, we choose to match $\tilde{\bm{A}}^{(i+1)}$ to $\tilde{\bm{A}}^{(i)}$ using degree profiles \cite{ding2020efficient} because of its probabilistic guarantees for exact recovery.

A degree profile is the empirical distribution of the degrees of a node's neighbors, which is computed for each node in each network.
Matching occurs by defining a distance matrix $Z^{(i, j)}$, where $Z_{uv}^{(i,j)}$ is the total variation distance between the degree profiles of nodes $u$ and $v$ in $\tilde{\bm{A}}^{(i)}$ and $\tilde{\bm{A}}^{(j)}$.
Ding et al. \cite{ding2020efficient} prove that for correlated Erd\H{o}s-R\'{e}nyi networks (with $\alpha = 0$), under certain conditions, the smallest $n$ entries of $Z$ recovers the exact permutation with high probability.

When the assumptions required in~\cite{ding2020efficient} do not hold, the authors recommend outputting an approximation, which can be accomplished by solving the following linear assignment problem:
\begin{equation*}
    \underset{S : \vert S \vert = n}\min \quad \sum_{(u,v)} \vert Z_{uv}^{(i,j)}\vert \quad \text{s.t } S = \{(u, v)\} \text{ is a permutation} \enskip .
\end{equation*}

We apply this procedure sequentially, which is summarized in Algorithm \ref{alg:matching}.
Note that Algorithm \ref{alg:matching} can be implemented in parallel and therefore has the same computational cost as Algorithm 1 in \cite{ding2020efficient} for $m = 2$, which is $O(nd^2 + n^2)$, where $d$ is the average degree.

\begin{algorithm}[htb!]
    \SetKwInOut{Input}{Input}
    \SetKwInOut{Output}{Output}
    \underline{Multi-graph match} $\big(\tilde{\bm{A}}^{(1)}, \ldots, \tilde{\bm{A}}^{(m)}\big)$\;
    \Input{$m$ unlabeled noisy networks, $\big\{\tilde{\bm{A}}^{(i)}\big\}_{i=1}^m$, on $n$ nodes}
    \Output{$m-1$ permutations $\hat{\pi}^{(1,2)}, \hat{\pi}^{(2,3)}, \cdots, \hat{\pi}^{(m-1,m)}$}
    \ForPar{$i \in [m-1]$}
    {
        Match $\tilde{\bm{A}}^{(i)}$ and $\tilde{\bm{A}}^{(i+1)}$ by degree profiles \cite{ding2020efficient} to obtain $\hat{\pi}^{(i, i+1)}$\;
    }
    \caption{Sequential pairwise graph matching procedure for aligning multiple unlabeled noisy networks.}
    \label{alg:matching}
\end{algorithm}

Given $\hat{\pi}^{(1, 2)}, \ldots, \hat{\pi}^{(m-1, m)}$ from Algorithm \ref{alg:matching}, let $\hat{\bm A}^{(i)}$ denote the adjacency matrix for $\tilde{\bm A}^{(i)}$ that has been realigned to $\tilde{\bm A}^{(1)} \equiv \hat{\bm A}^{(1)}$ for all $i = 2, \ldots, m$ based on the composition of the estimated permutations, i.e
\begin{align}\label{eq:align}
    \hat{A}^{(i)}_{uv} = \tilde{A}^{(i)}_{\hat{\pi}^{(i)}(u), \hat{\pi}^{(i)}(v)} \enskip ,
\end{align}
where $\hat{\pi}^{(i)} = \hat{\pi}^{(1, 2)} \circ \cdots \circ \hat{\pi}^{(i-1, i)}$ for $i = 2, \ldots, m$.
With this notation, we can compute our sample average in \eqref{eq:avg}.
Note that this average is conditional on the estimated permutations $\hat{\pi}^{(i-1, i)}$ for all $i = 2, \ldots, m$, and if these estimated permutations are correct, then this coincides with the sample Fr\`echet mean of $\tilde{\bm{A}}^{(1)}, \ldots, \tilde{\bm{A}}^{(m)}$ induced by the Frobenius distance, i.e
\begin{equation*}
    \widehat{\mu_m} := \underset{\bm A \in \mathcal{G}}{\arg\min} \frac{1}{m}\sum_{i=1}^m d_\text{F}^2\big(\bm A, \tilde{\bm{A}}^{(i)}\big) \enskip ,
\end{equation*}
where $\mathcal{G}$ is our space of networks and
\begin{equation}\label{eq:frobenius}
    d_\text{F}^2\big(\bm{A}^{(i)}, \bm{A}^{(j)}\big) = \frac{1}{n(n-1)}\sum_{u,v \in [n]}\big(A_{uv}^{(i)} - A_{uv}^{(j)}\big)^2 \enskip .
\end{equation}
However, note that we are computing a sample average and do not have a notion of population, since our samples are noisy versions of a single fixed network.
This is in contrast to the Fr\`echet mean estimated in \cite{kolaczyk2020averages}, which is an estimate of a population average over a given distribution of unlabeled networks.

In Section \ref{sec:theory}, we prove under certain conditions that $\hat{\bm A}$ in \eqref{eq:est} based on $\bar{\bm A}$ has nice properties in terms of recovering the latent parent network $\bm{A}$.
When these conditions do not hold, though, we can still improve this estimate by borrowing information across networks.
We describe two such methods next.

\subsection{Cleanup Procedure and Seeded Matching}\label{sec:mult}

In \cite{ding2020efficient}, the authors propose an iterative cleanup procedure, which solves the following sparse linear assignment problem for a prespecified number of iterations $t = 1, \ldots, T$:
\begin{align}\label{eq:clean}
    \hat{\pi}^{(i, j)}_t = \arg\max_\pi \langle \pi, \tilde{\bm{A}}^{(i)} \hat{\pi}^{(i, j)}_{t-1} \tilde{\bm{A}}^{(j)} \rangle \enskip ,
\end{align}
where $\hat{\pi}^{(i, j)}$ is initialized from the output of the graph matching procedure.

We can improve this given $m > 2$ networks by iteratively cleaning up pairs of networks, which we summarize in Algorithm \ref{alg:clean}.
This adds little computational overhead as the optimization in \eqref{eq:clean} is for sparse matrices ($\pi$ is a permutation matrix and thus $\tilde{\bm{A}}^{(i)} \hat{\pi}^{(i, j)}_{t-1} \tilde{\bm{A}}^{(j)}$ is sparse), which has highly efficient implementations using the Jonker-Volgenant algorithms \cite{jonker1987shortest}.

\begin{algorithm}[htb!]
    \SetKwInOut{Input}{Input}
    \SetKwInOut{Output}{Output}
    \underline{Multi-graph cleanup} $\big(\hat{\pi}^{(2)}, \ldots, \hat{\pi}^{(m)}\big)$\;
    \Input{$m-1$ permutations composed from output of Algorithm \ref{alg:matching}}
    \Output{$m-1$ permutations $\hat{\pi}^{(1)}, \cdots, \hat{\pi}^{(m)}$}
    \Repeat{convergence}
    {
        Sample $i < j$ randomly\;
        \For{$t = 1, \ldots, T$}
        {
            $\hat{\pi}^{(j)}_t = \arg\max_\pi \langle \pi, \hat{\bm{A}}^{(i)} \hat{\pi}^{(j)}_{t-1} \tilde{\bm{A}}^{(j)} \rangle$
        }
    }
    \caption{Cleanup procedure for multiple graph matching.}
    \label{alg:clean}
\end{algorithm}

Another approach to improving the sequential alignment procedure by incorporating the multiple samples is through seeded matching.
In graph matching, seeds are pairs of corresponding vertices between the two graphs that are prespecified prior to running the algorithm.
For matching via degree profiles, seeds are given as matched pairs
\begin{equation}\label{eq:seeds}
    S = \big\{(u,v) : a_u \geq \tau, b_v \geq \tau', Z_{uv} \leq \xi\big\} \enskip ,
\end{equation}
where $u$ and $v$ are nodes with high degrees $a_u$ and $b_v$, respectively, whose degree profiles are close in distance.
This is particularly useful for dense graphs, but still requires a sufficient number of seeds to have probabilistic guarantees.

We can expand the definition of seeds recursively to leverage the additional information when $m > 2$ as follows:
\begin{equation*}
    \begin{split}
        \tilde{S}^{(i, j)} = \bigcup_{k=1}^{m}  \ \big\{(u, v) : \exists \ w \text{ s.t } &(u, w) \in \tilde{S}^{(i, k)}, \\
        &(w, v) \in \tilde{S}^{(k, j)}\big\} \enskip ,
    \end{split}
\end{equation*}
where $\tilde{S}^{(i, i+1)}$ is initiated as in \eqref{eq:seeds} for $i = 1, \ldots, m-1$. 

Note that the degree profiles for every node in every network are already computed in Algorithm~\ref{alg:matching}.
Therefore, we only need to compute $\binom{m}{2} - (m-1) \approx \frac{m^2}{2}$  additional pairwise comparisons since $Z^{(i, i+1)}$ is already computed for $i = 1, \ldots, m-1$.
Furthermore, $\vert \tilde{S}^{(i, j)} \vert$ is bounded above by $(1-\frac{\log n}{nq})$-quantile of Binomial$(n-1, q)$, so the search time for the recursion is also bounded.

\section{Theoretical Results}\label{sec:theory}

In this section, we provide a probabilistic guarantee for when Algorithm \ref{alg:matching} exactly recovers the latent permutations.
Then, conditional on these latent permutations, we show that our estimate of the true underlying network from \eqref{eq:est} is asymptotically unbiased.

\subsection{Algorithm \ref{alg:matching} Performance}

We begin by showing that, with high probability, Algorithm \ref{alg:matching} can move us from an unlabeled network problem to a labeled network problem by exactly recovering the latent permutations.

\begin{theorem}\label{thm:rec}
    Let $\tilde{\bm{A}}^{(1)}, \ldots, \tilde{\bm{A}}^{(m)}$ be a collection of unlabeled networks sampled independently from \eqref{eq:noise} conditional on some fixed but unknown $\bm A \sim \text{ER}(n, q)$.
    Let $q$ and $s$ satisfy \eqref{eq:qs}.
    If $\sigma^2 = 1-s$ and $q \leq 1/12$ with
        \begin{equation*}
            \sigma \leq \frac{\sigma_0}{\log n}, \qquad L = L_0\log n, \qquad nq \geq C_0 \log^2n \enskip ,
        \end{equation*}
    for a sufficiently small constant $\sigma_0$ and sufficiently large constants  $L_0$ and $C_0$, then with probability $1 - O(1/n)$, Algorithm 1 outputs $\hat{\pi} = \pi^*$.
    It follows that Algorithm \ref{alg:matching} outputs
    \begin{equation*}
        \hat{\pi}^{(i,i+1)} = {\pi}^{*(i,i+1)} \enskip ,
    \end{equation*}
    for all $i \in [m-1]$ with probability $1 - O(m/n)$.
\end{theorem}

The proof of Theorem~\ref{thm:rec} is straightforward.
The first part is just Theorem 1 from \cite{ding2020efficient}, but with the recognition that if $\alpha > 0$, then, for any $\beta \in [0,1]$, $q$ and $s$ must be defined by \eqref{eq:qs}.
The conclusion follows from the fact that each permutation is recovered with probability $1 - O(1/n)$, hence the probability of recovering all $m-1$ permutations is $1 - O(m/n)$.
Therefore, we obtain exact recovery if $m = o(n)$, i.e so long as the network order grows faster than the sample size.

It may seem counterintuitive that the likelihood of recovery should decrease as our sample size increases.
This is due to the fact that our procedure takes a pairwise graph matching approach and so each comparison reduces the probability of the overall recovery.
However, in practice, we see that the methods in Section \ref{sec:mult} help mitigate the loss of accuracy as $m$ increases, which offsets the decrease in Theorem \ref{thm:rec}.

\subsection{Unbiased Recovery}

Here, we show that an unbiased estimator of the parent network can be defined as a function of the average of the aligned networks.

First, we have
\begin{equation}\label{eq:exp_avg}
    \begin{split}
        \mathbb{E}[\bar{A}_{uv}~|~\bm A] &= \mathbb{E}[\bar{A}_{uv}~|~\bm A, \hat\pi = \pi^*] \mathbb{P}(\hat\pi=\pi^*~|~\bm A) \\
        &\ \ \ + O(\mathbb{P}(\hat\pi \neq \pi^*~|~\bm A)) \enskip .
    \end{split}
\end{equation}
Note that the exact recovery in Theorem \ref{thm:rec} is marginal, i.e not conditional on $\bm A$.
We can overcome this with the following assumption.

\begin{assumption}\label{ass:whp}
    \begin{equation*}
        \mathbb{P}(\bm A~|~\hat\pi = \pi^*) \geq \mathbb{P}(\bm A) \enskip .
    \end{equation*}
\end{assumption}

Assumption \ref{ass:whp} implies that parent networks have higher likelihoods when matching via degree profiles successfully recovers the latent permutation of two noisy replicates.  This assumption can be expected to be difficult to verify in practice, given that unsurprisingly the conditional likelihood is intractable.
Nevertheless, this is intuitively reasonable and is also corroborated by simulation.

It follows from Assumption \ref{ass:whp} that $\hat\pi=\pi^*$ with high probability conditional on $\bm A$:
\begin{align*}
    \mathbb{P}(\hat\pi=\pi^*~|~\bm A)
    &= \frac{\mathbb{P}(\bm A~|~\hat\pi=\pi^*)\mathbb{P}(\hat\pi=\pi^*)}{\mathbb{P}(\bm A)} \\
    &\geq \frac{\mathbb{P}(\bm A)\mathbb{P}(\hat\pi=\pi^*)}{\mathbb{P}(\bm A)} \\
    &= \mathbb{P}(\hat\pi=\pi^*) \\
    &= 1~\text{as}~n \to \infty \enskip .
\end{align*}

From the definition of the aligned average in \eqref{eq:avg}, we have
\begin{align*}
    \mathbb{E}[\bar{A}_{uv}~|~\bm A, \hat\pi = \pi^*]
    &= \frac{1}{m}\sum_{i=1}^m\mathbb{E}[\tilde{A}^{(i)}_{uv}~|~A_{uv}, \hat\pi = \pi^*] \\
    &= \frac{1}{m}\sum_{i=1}^m (1-\beta) \cdot {A_{uv}} + \alpha \cdot (1- A_{uv}) \\
    &= \begin{cases}
        1 - \beta & A_{uv} = 1 \\
        \alpha & A_{uv} = 0
    \end{cases} \enskip .
\end{align*}

Therefore, for large $n$, we have unbiased recovery of $\bm A$ when there is adequate separation between the noise rates.

\begin{assumption}\label{ass:noise}
    \begin{equation*}
        1 - \beta > \alpha \enskip .
    \end{equation*}
\end{assumption}

Assumption \ref{ass:noise} stipulates that there is enough signal compared to the noise in \eqref{eq:noise} and is much weaker than other methods in the noisy network literature that require the noise rates to be known or estimated.
We do not need to know $\alpha$ or $\beta$, but rather an appropriate threshold $w$ that separates true edges from spurious ones.
In practice, $w$ is also unknown, but we find that an elbow method works well empirically for choosing a break $\hat{w}$ in estimated edge weights for our estimate in \eqref{eq:est}.

Given $w$ or an unbiased estimate $\hat{w}$, it follows that, with high probability, we have an unbiased estimate of $\bm A$, i.e for all $u, v$, 
\begin{equation*}
    \mathbb{E}[\hat{A}_{uv}~|~\bm A] \to A_{uv}~\text{as}~n \to \infty \enskip .
\end{equation*}
Furthermore, the assumption of independent noise implies $\hat{\bm A}^{(1)}, \ldots, \hat{\bm A}^{(m)}$ are independent conditional on $\bm A$.
Therefore, the central limit theorem from \cite{ginestet2017hypothesis} for labeled networks holds for our sample average $\bar{\bm A}$ in \eqref{eq:avg} in case confidence intervals are desired.

\section{Simulations}\label{sec:simulation}

In this section, we present the results from a simulation study.
All of the code for our algorithms and reproducing our simulation is available at \href{https://github.com/KolaczykResearch/NetworkRecovery}{https://github.com/KolaczykResearch/NetworkRecovery}.

We mimic the setup in Figure 5 of \cite{ding2020efficient} by varying $n$ and $\beta$ while letting $p = \log^2(n)/n$.
However, we also vary $m$, as well as introducing $\alpha > 0$, which we do by setting $\alpha$ to satisfy edge unbiasededness, i.e
\begin{equation*}
    \alpha \cdot |E^c| = \beta \cdot |E| \enskip .
\end{equation*}
The assumption of edge unbiasedness guarantees the expected number of edges in the noisy observation is equal to the number of edges, $|E|$, in the true network.
This also ensures that Assumption \ref{ass:noise} is satisfied for large $n$ because $\alpha~=~\Theta(\beta~\cdot~\log^2(n)~/~n)$ and hence $\hat{w} = .5$ can be used an appropriate threshold.

\begin{figure}[!htb]
    \centering
    \includegraphics[width=.5\textwidth]{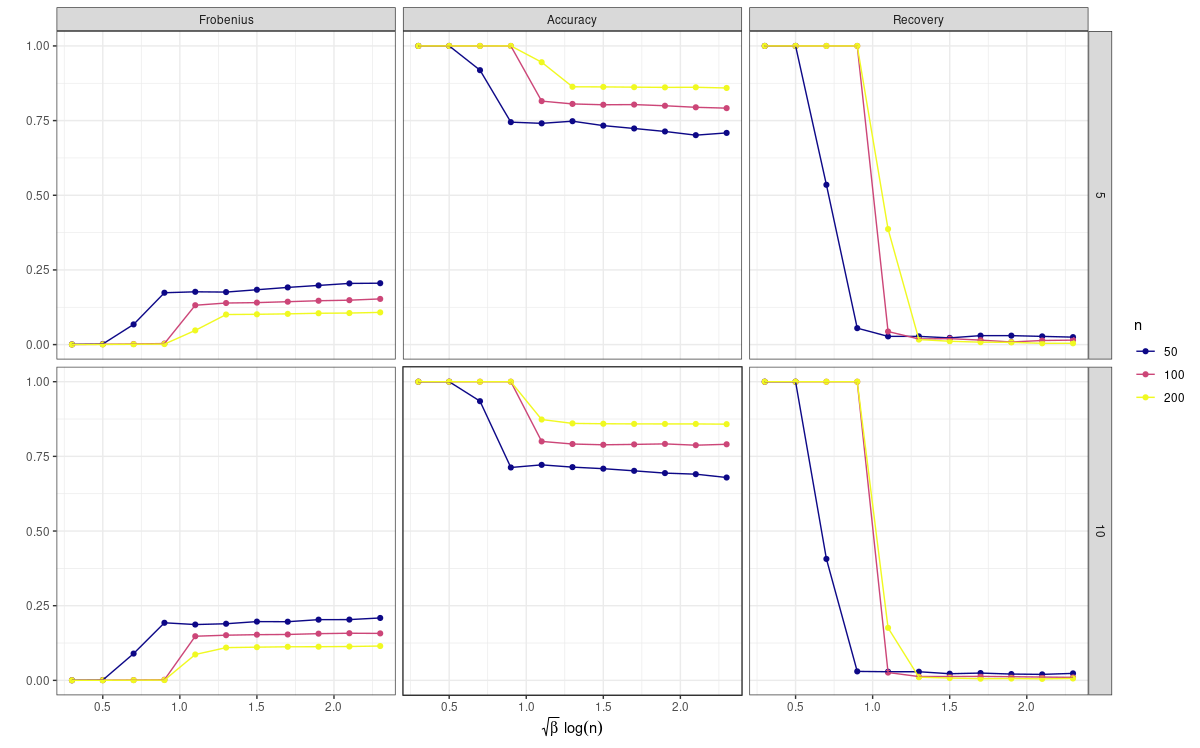}
    \caption{Results from the simulation. The rows represent the number of networks $m$ and the colors represent the number of nodes $n$. The columns display different accuracy measurements, where the first and second show the Frobenius distance and accuracy, respectively, between the aligned average from \eqref{eq:avg} and the estimate from \eqref{eq:est} compared to the true underlying network. The third column shows the accuracy of multi-graph matching with multi-graph cleanup from Algorithms~\ref{alg:matching} and \ref{alg:clean}.}
    \label{fig:clt_sim}
\end{figure}

For each of the simulation parameters, we compute the fraction of correctly matched pairs, the squared Frobenius distance in \eqref{eq:frobenius} between $\bar{\bm A}$ and $\bm A$, and the fraction of correctly identified edges and non-edges between $\hat{\bm A}$ and $\bm A$.
These are referred to as Recovery, Frobenius, and Accuracy, respectively.
For each of the settings, we run 10 independent trials and report the median for each of these measures.
The results are given in Figure \ref{fig:clt_sim}.

The rightmost column for Recovery parallels the findings in Figure 5 of \cite{ding2020efficient}.
In particular, we have the same results even with $m > 2$ and $\alpha > 0$, which is what we would expect from Theorem \ref{thm:rec}.
Furthermore, Frobenius is low and Accuracy is high when Recovery is high, which is also what we would expect.
The plateaus in Frobenius and Accuracy as $\sqrt{\beta} \log(n)$ increases correspond to the fact that $p = \log^2(n) / n$, i.e as $n$ increases, the density decreases and thus most potential edges are (correctly) predicted to be absent.
Finally, it is worth noting that the sharp decline in the results around $\sqrt{\beta} \log(n) \geq 0.7 $ has been described as ``all-or-nothing recovery" and is being extensively studied in the case of $\alpha = 0$ \cite{hall2020partial, wu2021settling, ganassali2021impossibility}.

\section{Conclusion}\label{sec:conclusion}

\subsection{Summary}

In this work, we turned our attention to the underdeveloped area of multiple unlabeled networks.
In order to overcome the additional combinatorial challenge arising from node permutations, we focused on networks arising from a `signal-plus-noise' model in which the observed networks are unlabeled noisy samples from a true underlying network.
This setting allowed us to leverage results from the graph matching literature and we developed a procedure that sequentially aligns the unlabeled networks by recovering the latent permutations.
In doing so, we moved the problem to the labeled setting with high probability, which yielded an unbiased estimator of the underlying network.
This solution is computationally efficient and will hopefully serve as a starting point for future analysis of multiple unlabeled networks.

\subsection{Future Work}

The theoretical guarantee of the performance of graph matching via degree profiles relies on the distribution of the node degrees in each network.
Originally, the authors only considered the correlated Erd\H{o}s-R\'{e}nyi model given in Definition \ref{def:corrER}.
In this setting, the node degrees are binomial random variables and hence exhibit nice tail bounds.
As we verified, the theory holds in the generative framework even when $\alpha$, the type I noise rate, is positive.
However, we still only considered the case when the true underlying network $\bm A$ is Erd\H{o}s-R\'{e}nyi, hence the edge probabilities for the underlying network $\bm A$ are homogeneous.
It is worth pursuing whether the results hold if the true underlying network $\bm A$ is from an inhomogeneous Erd\H{o}s-R\'{e}nyi in which the true edges are still drawn from Bernoulli distributions but with possibly different probabilities.

Another interesting question is whether the results hold in the case where the number of nodes is random.
We could achieve this by assuming a Poisson distribution on the network order, thus inducing a hierarchical model.


\bibliographystyle{IEEEtran}
\bibliography{refs.bib}

\end{document}